\documentclass[11pt, a4paper,superscriptaddress,aps,nofootinbib,  ]{revtex4}
\usepackage{amsmath,amssymb,graphicx}
\makeatletter
\usepackage[active]{srcltx}
\usepackage{graphicx,color}
\usepackage{changebar}
\usepackage{hyperref}
\hypersetup{
	colorlinks=true,
	urlcolor= blue,
	citecolor=blue,
	linkcolor= blue,
	bookmarks=true,
	bookmarksopen=false,
}
\usepackage[T1]{fontenc}
\usepackage{ae}
\usepackage[ansinew]{inputenc}
\usepackage{amsmath}
\usepackage{braket}
\usepackage{mathtools}
\usepackage{slashed}
\usepackage{empheq}
\usepackage{tikz}
\usepackage{multirow}
\usepackage[caption = false]{subfig}

\usepackage{graphicx}
\usepackage{amsfonts}
\usepackage{amsmath}
\usetikzlibrary{arrows} 
\usetikzlibrary{decorations.markings}

\begin{document}
\title{Graphene-based topological insulator in the presence of a disclination submitted to a uniform magnetic field}

\author{J. R. S. Oliveira}
\email[]{jardson.ricardo@gmail.com}
\affiliation{Departamento de F\'{\i}sica, Universidade Federal da 
	Para\'{\i}ba,\\
	Caixa Postal 5008, 58051-970, Jo\~ao Pessoa, Para\'{\i}ba, Brazil}

\author{G. Q. Garcia}
\email[]{gqgarcia99@gmail.com}
\affiliation{Unidade Acad\^emica de F\'isica, Universidade Federal da Campina Grande, 58109-970 Campina Grande, PB, Brazil}

\author{P. J. Porf\'{i}rio}\email[]{pporfirio89@gmail.com}
\affiliation{Departamento de F\'{\i}sica, Universidade Federal da 
	Para\'{\i}ba,\\
	Caixa Postal 5008, 58051-970, Jo\~ao Pessoa, Para\'{\i}ba, Brazil}

\author{C. Furtado}
\email[]{furtado@fisica.ufpb.br}
\affiliation{Departamento de F\'{\i}sica, Universidade Federal da 
	Para\'{\i}ba,\\
	Caixa Postal 5008, 58051-970, Jo\~ao Pessoa, Para\'{\i}ba, Brazil}

\begin{abstract}
In this paper we investigate quasiparticles in graphene-based topological insulator with a wedge disclination in the presence of a uniform magnetic field. In particular, we consider a massive spinless model. The Landau levels are analytically found for the described system. We show that the Landau levels are lifted for a specific range of the parameter $\nu$. In addition, we also show that, for appropriate parameter values, one finds gapless and gapped states.  
\end{abstract}

\maketitle
 
 \section{INTRODUCTION}
 \label{Int.}

 Graphene has a large number of important properties widely studied nowadays. Among them there are some effects of the theory of relativity and quantum electrodynamics in low dimensions that can be mimicked by laboratory experiments \cite{gein,been,castro}. Graphene is a honeycomb lattice made up of a set of carbon atoms. The carbon atoms are distributed in each of hexagonal vertices of the structure with each unit cell having two carbon atoms. Indeed, the honeycomb lattice can be considered as a superposition of two $\mathcal{A/B}$ triangular sublattices. The Dirac electrons in graphene occur in two degenerate families that correspond to the presence of two different valleys in the structure of bands - a phenomenon known as doublet fermions. Due to degeneracy, in many cases, it is difficult to observe the physics of a simple valley in experiments, because of the two valleys have equal and opposite contributions to the measurable quantities. One way to break down this degeneracy is to produce a fictitious magnetic field in a simple valley by a defect in the crystalline lattice. The field has an opposite sign in the other valley, so it raises the degeneracy. The broken valley degeneracy leads to some interesting contributions of quantum oscillations.

Haldane \cite{haldane} has considered a model for a honeycomb lattice without spin to study the quantum Hall effect, without magnetic field (no Landau levels), and maintaining the translation symmetry of the lattice. In this model, Haldane \cite{haldane} advocated that the quantum Hall effect should be placed in a broader context in which time-reversal symmetry is broken down. This system investigated by Haldene is the first model of Chern insulator (CI), also known as Haldane insulator, and it is a quantum Hall anomalous phase. Despite being a simplified model, Kane and Mele employed it to describe Hall Effect of Quantum Spin \cite{prl:kane}. It is well known that the classification of different phases of matter is indeed a topic of great interest in several areas of physics. In turn, some phase scenes can be interpreted and analysed using a Landau theory. This is the type of phase transition that characterizes states whose symmetries have been broken. So that the electronic states operated by the Quantum Hall Effect (QHE) there will be no broken symmetries, but they define a topological phase  with respect to some fundamental properties like the quantized value of Hall Conductance. For strong magnetic fields, this conductivity is quantized due to the Landau levels and the number of gapless modes at the edges of the sample are insensitive to soft changes in the parameters of the material, and do not change for a system that does not undergo a quantum phase transition. 

Kane and Mele \cite{prl:kane} did investigations on the effects of the intrinsic($\Delta$) and Rashba type($\lambda_{R}$) spin-orbit coupling (SOC) in a monolayer of graphene and have showed that the spin-orbit interaction converts the ideal two-dimensional semi-metallic state of graphene to a quantum spin Hall insulator. Thus, creating a new electronic state of the matter that is gapped in the bulk,  but with topologically protected edge states near the boundary of the sample\cite{RMP:kane}, this new class of material was denominated topological insulators. In Ref.\cite{pan}, Pan {\it et al.} performed experiments using techniques  angle-resolved photoemission spectroscopy in  $Bi2Se3 $ and was obtained that this material have behaved as a  topological insulator. Also, Roushan {\it et al.} \cite{rous} using the same experimental technique used in Ref.~\cite{pan} have studied the $Bi12xSbx$ alloys and concluded that this material had  topological insulator  characteristics, such as, a large energy gap and single-surface Dirac cone  associated with  its topologically protected  state\cite{dubo}. In a recent work, De Martino {\it et al.}\cite{demart} have analyzed the electronic properties of graphene monolayer in the presence of a magnetic and pseudomagnetic field taking into account SOC interactions,  they have found the Landau levels  considering the intrinsic and Rashba interactions. The Landau levels for a series of  gapped Dirac materials, i.e., silicene, germanene, etc,   with intrinsic Rashba interaction was investigated by Tsaran and  Sharapov\cite{tsar}. The Landau levels in graphene layer with Rashba coupling was investigated in Ref.\cite{rash}, where was observed that two zero modes emerge in the eigenvalues of energy. A theoretical investigation of the Landau levels in bilayer graphene was made by Mireles and Schliemann\cite{mire} in the limit of low energy in the presence of transverse magnetic field and a Rashba SOC. The  bound states of energy in Graphene-based topological insulator have been studied by De Martino {\it et al.}\cite{cris} and they have obtained the spectrum of energy for Kane-Mele Hamiltonian
in the presence of an attractive potential and considered intrinsic and Rashba SOC.  In recent years, the study of the influence of topological defects in electronic properties of graphene has been widely investigated by several authors~\cite{rmp:voz,epl:voz,prl:lam,pla:car,KnuteClaudio1,knut,carvalho}. Recently,  R\"uegg and Lin~\cite{rueg:prb} have investigated the bound states induced by disclinations  in  Graphene-based  topological insulators. R\"uegg {\it et al.} have investigated the  Haldane honeycomb lattice model\cite{hald} on topologically spherical nanosurfaces: tetrahedron, the octahedron and the icosahedron, and  conclude that each corner of such structures carries
a non-trivial bound state, they are named nanostructures of topological fullerenes. Choudhari and  Deo \cite{epl:niv} have studied  the influence of a disclination in a graphene layer where was considered non-trivial SOC, intrinsic and Rashba  couplings, and was found  the eigenvalues and eigenfunctions for a modified Kane-Mele Hamiltonian obtained for a graphene layer with disclination, for a case of intrinsic SOC  the problem was solved exactly . In this contribution we consider the modified Kane-Mele Hamiltonian for disclinated graphene layer considering the intrinsic SOC in the presence of magnetic flux and a uniform magnetic field in $ z$-direction. We obtain analytically the eigenfunctions and eigenvalues of energy  for Landau levels in this geometry. We investigate the influence of the parameter that characterizes the disclination and discuss the physical implications.

The structure of this paper is organized as follows. In section \ref{sec2}, we describe the main features of the Kane-Mele Hamiltonian for a graphene layer with a mass term. Afterwards, in section \ref{sec3}, we obtain the energy spectrum for the graphene layer in the presence of wedge disclinations without spin and discuss their main physical consequences. Ultimately, in section \ref{End}, we provide a summary and conclusions of our main results.
  
\section{model for a graphene layer}\label{sec2}

In this section we briefly discuss the general model we will use throughout this work.
To start with, let us consider the Kane-Mele model with an additional mass term. Kane and Mele developed a model for graphene \cite{prl:kane} using the tight-biding model taking on the possibility of the electron hopping to the second neighbors. In that case, the modified Kane-Mele Hamiltonian looks like
\begin{eqnarray}
\mathcal{H} = \sum_{<ij>}t c^{\dagger}_{i \alpha} c_{j \alpha} + \sum_{<<i j>> \alpha \beta} i t_{2} \nu_{ij}s^{z}_{\alpha \beta} c^{\dagger}_{i \alpha} c_{j \beta} +  m c^{\dagger}_{i \alpha} c_{i \alpha}.
\label{2.1}
\end{eqnarray}
Note that, the first term in equation (\ref{2.1}) is the usual tight-biding Hamiltonian for graphene considering the interaction with the first neighbors in the lattice. From it, we can see that the $t$-parameter gives us the probability of one electron present in the site $\mathcal{A/B}$ to hop for the first neighbors, i.e., the site $\mathcal{B/A}$. The second term describes the second neighbors interaction with spin-dependent amplitude. The parameter $t_{2}$ characterizes the second neighbors hopping amplitude. In addition, $s^{z}_{\alpha \beta}$ is the spin operator for an electron with spin lying in $z$-direction. The parameter $\nu_{ij}$ depends on the orientation of the two next nearest neighbor bonds, say $\vec{d}_{1}$ and $\vec{d}_{2}$. For convenience, let us choose the following convention: anticlockwise - defined as positive hopping, i.e., hopping from the site $j$ to $i$ corresponds to $\nu_{ji}=+1$ whilst clockwise - defined as negative hopping, i.e., corresponds to $\nu_{ji}=-1$. Finally, the last term is the mass one, which is the new term we are interested in. 

At this moment, we are able to write the tight-binding Hamiltonian in the low-energy limit for an electron with spin and in the presence of a mass term:
\begin{eqnarray}
\mathcal{H} = v_{f}\hbar(\tau_{z}\sigma_{x}k_{x} + \sigma_{y}k_{y}) + \Delta_{soc} \tau_{z}\sigma_{z} s_{z} + m\tau_{z}\sigma_{z}.
\label{2.2}
\end{eqnarray} 
It is important to point out that the equation (\ref{2.2}) has three sets of Pauli matrices: $\tau_{i}$, $\sigma_{i}$ and $s_{i}$ acting in their respective spaces, namely: valley, sublattice, and spin spaces. Here, we are using the particular notation where $\tau_{z}=\pm 1$ for two valleys $K$($K'$) in Brillouin zone, $\sigma_{z}=\pm 1$ for sublattice $\mathcal{A/B}$, and $s_{z}=\pm 1$ for spin up or down of the electron. Therefore each valley ($K$ or $K'$) in Brillouin zone has four degrees of freedom, and the Hamiltonian $\mathcal{H}$ acts on a eight-component spinor: 
\begin{equation}
\Psi= [(\psi_{\mathcal{A}\uparrow}\psi_{\mathcal{A}\downarrow}\psi_{\mathcal{B}\uparrow}\psi_{\mathcal{B}\downarrow}),      (\psi_{\mathcal{A}'\uparrow}\psi_{\mathcal{A}'\downarrow}\psi_{\mathcal{B}'\uparrow}\psi_{\mathcal{B}'\downarrow})]^{T},
\nonumber
\end{equation}
where $\mathcal{A}$ and $\mathcal{B}$ label the sublattices in the valley $K$, and $\mathcal{A}'$ and $\mathcal{B}'$ label the sublattices in the valley $K'$. 

\section{Energy levels for disclinated graphene with a mass term}\label{sec3}

So far we have introduced the modified Kane-Mele Hamiltonian for a simple graphene layer, however the aim in this present study is to exploit the graphene in the presence of a topological defect, more precisely, a disclination. Topological defects are quite common in the preparation of real solid crystals. In particular, disclinations are related to the arising of pentagons or heptagons in the honeycomb lattice, as a consequence, it has a failure in the atoms distribution on two sublattices $\mathcal{A/B}$. The presence of topological defects in the crystalline lattice was explained by Volterra in \cite{volterra} using the ``cut and glue" processes. For example, in graphene lattice, the presence of pentagons and heptagons is due to the removal (or the insertion) of an angular sector of the flat lattice by cutting and gluing its edges to form a lattice with three bonds per atom. This angular sector is an integer multiple of $\dfrac{\pi}{3}$, i.e., $\lambda=N\dfrac{\pi}{3}$, with $-6<N<6$. As expected this procedure provides a non-trivial curvature to the resulting lattice. Indeed, removing an angular sector ($N>0$) generates a positive curvature  and it sets up the conical shape of the graphene, while the insertion of an angular sector ($N<0$) engenders a negative curvature. Thenceforth, for the sake of convenience, we shall use the angular deficit parameter $\alpha_{N}$ to characterize the disclination, it is defined by $\alpha_{N}=\left(1-\frac{N}{6}\right)$.

 We now focus our attention on positive disclinations, i.e., the graphene with a conical shape. It is worth stressing that the presence of pentagon deffects ($N=1$) in graphene lattice has the connection between two sites belonging the same sublattice, such that if we rotate the spinor around the disclination core, it leads to a non-trivial holonomy. The best manner to explore the physical properties of the holonomy is to split it into two singular contributions, namely: the first one is related to the parallel transport of spinor around the apex of the cone in a closed path, as a result, it provides the variation of the local reference frame along the path \cite{prl:lam,pla:car,rueg}. Its explicit form is given by $U_{s}(\phi)=e^{i\frac{\phi}{2}\tau_{z}\sigma_{z}s_{0}}$. It has been shown in \cite{rueg,epl:niv}, the holonomy $U_{s}(\phi)$ is responsible for transforming $\Psi$ to a  corotating spinor. The second contribution for the holonomy is a Aharonov-Bohm-like contribution \cite{pr:ahan, prl:lam} and is given by $V_{ns}(\theta)=e^{i\frac{n\theta}{4}\tau_{y}\sigma_{y}s_{0}}$. As pointed out in \cite{jrso:annals}, it introduces a matrix-valued gauge field in the Hamiltonian. Therefore, the holonomy imposes the following boundary condition for the Dirac spinor:
\begin{eqnarray}\label{3.1}
\Psi(\theta=2\pi)=U_{s}(\phi)V_{ns}(\theta)\Psi(\theta=0),
\end{eqnarray}
where we have transformed the angle $\phi$ in the unfolded plane into the polar angle $\theta=\frac{\phi}{\alpha_{N}}$, with $0<\theta<2\pi$, see \cite{jrso:annals} for a detail discussion.

To proceed further, let us also introduce two extra fields in this model through the minimal coupling procedure in the modified Kane-Mele Hamiltonian. The first is a field related to the holonomy (\ref{3.1}), and it is similar to the Aharonov-Bohm flux \cite{pr:ahan} defined in a conical space. Actually, this field plays a role of a potential vector of the ``fictitious'' magnetic flux generated by the disclination source. Such a procedure is well-known and has been discussed in details in \cite{rueg,epl:niv}. Following these references, the potential vector is given by
\begin{eqnarray}\label{3.2}
\vec{A}=\frac{\Phi}{r\alpha_{N}\Phi_{0}} \hat{\phi},
\end{eqnarray}
where $\Phi_{0}=h/e$ is the quantum magnetic flux. 

In regard the second field, we implement it as being an external magnetic field in $z$-direction, we mean, $\vec{B} = B_{0}\hat{z}$, defined in conical space. Note that, this configuration of magnetic field is known as {\it symmetric gauge} and its associated potential vector has been studied in conical space in\cite{pla:bruno}. From this configuration we can generate the potential vector in the continuum space with the presence of disclination as follow, 
\begin{eqnarray}\label{3.3}
\vec{A_{r}}= \frac{B_{0}r}{2}\hat{\phi}.
\end{eqnarray} 

Having implemented these ingredients we are now ready to find the energy spectrum. In particular, we will concentrate our attention on the effects of the disclination and the mass term, so, from now on, we disregard SOC interactions, $\Delta_{soc}=0$. In this case, the modified Kane-Mele Hamiltonian reduces to the Haldane Hamiltonian with a mass term. It is noteworthy that now it suffices two degrees of freedom in order to describe an irreducible spin representation for this system, i.e., $\Psi=[(\psi_{\mathcal{A}}\psi_{\mathcal{B}}),(\psi_{\mathcal{A}^{\prime}}\psi_{\mathcal{B}^{\prime}})]^{T}$. The next step is to solve the eigenvalue equation for the Haldane Hamiltonian, $H_{MH}\Psi=\epsilon\Psi$, for a disclinated graphene with a mass term. In terms of the coordinates $(r,\theta)$, the momentum operator in equation (\ref{2.2}) is defined by $k_{r}=-i\frac{\partial}{\partial r}$ and $k_{\theta}=\frac{-i}{r\alpha_{N}}\frac{\partial} {\partial\theta}$. By introducing those two extra fields (\ref{3.2}) and (\ref{3.3}), then the resulting Hamiltonian reads
\begin{eqnarray}\label{3.4}
H_{MH}= v_{f}\hbar \bigg[\tau_{z}\sigma_{x}k_{r} + \sigma{y}\bigg(k_{\theta} + \frac{\Phi}{r\alpha_{N}\Phi_{0}} +  \frac{e B_{0}r}{2\hbar}\bigg) \bigg] + m\tau_{z}\sigma_{z}.
\end{eqnarray}
The holonomy (\ref{3.1}) lead us to the gauge transformation in (\ref{3.4}) in the following way $H_{1MH} =U_{s}^{\dagger}(\phi) V^{\dagger}_{ns}(\theta)H_{MH}U_{s}(\phi) V_{ns}(\theta)$, and thereby the transformed  Hamiltonian for disclinated lattice is
\begin{eqnarray}\label{3.5}
H_{1MH} &&= \left[ k_{r}- \frac{i}{2r} \right]\tau_{z}\sigma_{x} + \left[k_{\theta}+ \frac{\Phi}{r\alpha_{N}\Phi_{0}} 
+  \frac{e B_{0}r}{2} + \frac{n}{4r\alpha_{N}} \right]\sigma_{y}+ m\tau_{z}\sigma_{z},
\end{eqnarray}
where we have used natural units $\hbar= v_{f}=1$.

Assuming the azimuthal symmetry, it suggests introducing the {\it ansatz} $\Psi(r,\theta)=e^{ij\theta}X(r)$, where $j$ is a half integer number $j = \pm1/2, \pm3/2,...$ Using this {\it ansatz}, we obtain
\begin{eqnarray}\label{3.6}
H'_{1MH}X(r)= \left[ \left(k_{r}- \frac{i}{2r} \right)\tau_{z}\sigma_{x} + \left(\frac{\nu_{\tau}}{r} + \omega r \right)\sigma_{y} + m\tau_{z}\sigma_{z}\right]X(r) = \epsilon X(r),
\end{eqnarray} 
where we have defined the frequency as $\omega = \frac{e B_{0}}{2}$, and $\tau=\pm 1$ for two emergent valleys. In addition, we have also used the shorthand notation:
\begin{eqnarray}
\nu_{\tau}=\frac{j+\frac{\Phi}{\Phi_{0}}+\frac{N\tau}{4}}{\alpha_{N}}.
\label{3.7}
\end{eqnarray}

The eigenfunction $X(r)$ belongs to the spinor space and then can be represented as a column matrix with two components. As a result, we obtain a set of two coupled differential equations corresponding to the first valley $K$. The sublattices $\mathcal{A/B}$ play a role of pseudospin. Then, the two differential equations can be written as:
\begin{eqnarray}
\label{acoplada1}-i{\frac {d}{dr}}{\it X_{B}} \left( r \right) -{\frac {i \left( 1/2+{\it 
\nu_{+}} \right) {\it X_{B}} \left( r \right) }{r}}-i{\it \omega}\,r{\it X_{B}}
 \left( r \right) - \left( \epsilon -m \right) {\it X_{A}} \left( r
 \right) =0;\\
\label{acoplada2}-i{\frac {d}{dr}}{\it X_{A}} \left( r \right) +{\frac {i \left( -1/2+{
\it \nu_{+}} \right) {\it  X_{A}} \left( r \right) }{r}}+i{\it \omega}\,r{\it  X_{A}}
 \left( r \right) - \left( \epsilon +m \right) {\it  X_{B}} \left( r
 \right) =0.
\end{eqnarray} 
Proceeding in a similar way to the first valley, one finds the set of coupled differential equations describing the second valley $K'$. Therefore, we can write down
\begin{eqnarray}
\label{acoplada1'}i{\frac {d}{dr}}{\it X_{B'}} \left( r \right) +{\frac {i \left( 1/2-{\it 
\nu_{-}} \right) {\it X_{B'}} \left( r \right) }{r}}-i\omega\,r{\it X_{B'}}
 \left( r \right) - \left( \epsilon +m \right) {\it X_{A'}} \left( r
 \right) =0;\\
\label{acoplada2'}i{\frac {d}{dr}}{\it X_{A'}} \left( r \right) +{\frac {i \left( 1/2+{\it 
\nu_{-}} \right) {\it X_{A'}} \left( r \right) }{r}}+i{\it \omega}\,r{\it X_{A'}}
 \left( r \right) - \left( \epsilon -m \right) {\it X_{B'}} \left( r
 \right) =0,
\end{eqnarray} 
where the prime in the former set of equations indicates we are referring to the valley $K'$.

Note that for each set of differential equations above, one for the valley $K$ and the other for the valley $K'$, we have one pair of first order coupled differential equation. In order to decouple them, we substitute Eq.(\ref{acoplada1}) in Eq.(\ref{acoplada2}) to find a second order decoupled differential equation for $X_{B}$. Similarly, substituting Eq.(\ref{acoplada2}) in Eq. (\ref{acoplada1}) we find a second order decoupled differential equation for $X_{A}$. Accordingly, we find
\begin{equation}
\begin{split}
&\frac{d^{2}X_{A}}{d r ^{2}} +  \frac{1}{r} \frac{d X_{A}}{d r} - \left[ \frac{\chi^{2}_{A}}{r^2} + \omega^{2} r^{2} - \beta_{A} \right]X_{A}  = 0; \\
&\frac{d^{2}X_{B}}{d r ^{2}} +  \frac{1}{r}  \frac{d X_{B}}{d r} - \left[ \frac{\chi^{2}_{B}}{r^2} + \omega^{2} r^{2} - \beta_{B} \right]X_{B}  = 0,
\label{eqsdesacopladas1}
\end{split}
\end{equation}
The same procedure should be adopted to find the decoupled differential equations for the valley $K^{\prime}$, thus arriving at
\begin{equation}
\begin{split}
&\frac{d^{2}X_{A'}}{d r ^{2}} +  \frac{1}{r} \frac{d X_{A'}}{d r} - \left[\frac{\chi^{2}_{A'}}{r^2} + \omega^{2} r^{2} - \beta_{A'} \right]X_{A'}  = 0; \\
&\frac{d^{2}X_{B'}}{d r ^{2}} +  \frac{1}{r}  \frac{d X_{B'}}{d r} - \left[\frac{\chi^{2}_{B'}}{r^2} + \omega^{2} r^{2} - \beta_{B'}  \right]X_{B'}  = 0.
\label{eqsdesacopladas2}
\end{split}
\end{equation}
where we defined the parameters as follow, 
\begin{eqnarray}
\chi_{A} &=& \nu_{+}-\frac{1}{2};\\
\chi_{B} &=& \nu_{+}+\frac{1}{2};\\
\chi_{A'} &=& \nu_{-}+\frac{1}{2};\\
\chi_{B'} &=& \nu_{-}-\frac{1}{2},
\end{eqnarray}
and,
\begin{eqnarray}
\beta_{A} &=& \left(\epsilon^{2} - m^{2}\right) - 2\omega\left(\nu_{+} + \frac{1}{2}\right);\\
\beta_{B} &=& \left(\epsilon^{2} - m^{2}\right) - 2\omega\left(\nu_{+} - \frac{1}{2}\right);\\
\beta_{A'} &=& \left(\epsilon^{2} - m^{2}\right) - 2\omega\left(\nu_{-} - \frac{1}{2}\right);\\  
\beta_{B'} &=& \left(\epsilon^{2} - m^{2}\right) - 2\omega\left(\nu_{-} + \frac{1}{2}\right).
\end{eqnarray}

To gain any further insight on the solutions of Eqs. (\ref{eqsdesacopladas1}) and (\ref{eqsdesacopladas2}), let us first examine their asymptotic behavior near the singular points: $r \to 0 $ and $r \to \infty$. Before we keep going, it is convenient to take the following coordinate transformation: $\rho = \omega r^{2}$. By doing so, we can write them in a more compact way,
\begin{eqnarray}\label{mudvar}
\rho\frac{d^{2}X_{\gamma}}{d\rho^{2}} + \frac{dX_{\gamma}}{d\rho} + \left[ \frac{\beta_{\gamma}}{4\omega} - \frac{1}{2} - \frac{\chi^{2}_{\gamma}}{4\rho}\right]X_{\gamma} = 0,
\end{eqnarray}
where $\gamma = A, A', B$ or $B'$ labels the components of spinor. Furthermore, proceeding with the following redefinition: 
\begin{eqnarray}\label{assint}
&&X_{\gamma}(\rho)= e^{-\frac{\rho}{2}}\rho^{\frac{|\chi_{\gamma}|}{2}}F_{\gamma}(\rho).
\end{eqnarray}
Note that the equation (\ref{assint}) are the asymptotic solutions from each equation represented by equation (\ref{mudvar}). The set of functions $F_{\gamma}(\rho)$ are the confluent hypergeometric functions, and they are solutions of following confluent hypergeometric differential equations:
\begin{equation}
 \rho\frac{d^{2}F_{\gamma}(\rho)}{d\rho^{2}} + \left(\vert\chi\vert+1-\rho\right)\frac{dF_{\gamma}(\rho)}{d\rho}+ \left[\frac{\beta_{\gamma}}{4\omega}-\frac{1}{2}-\frac{\vert\chi_{\gamma}\vert}{2}\right]F_{\gamma}(\rho) = 0
\end{equation}

Typically, we must impose boundary conditions in order to get regular solutions at the singular points: $\rho=0$ and $\rho=\infty$. Using the special properties of hypergeometric functions \cite{book}, the regularity conditions are attained by requiring that the first argument of the hypergeometric function $\,_{1}F_{1}(a,b,\rho)$ be identified as a non-positive integer number, i.e., $a=-n$, thus $n=0,1,2...$. Such a condition reduces the hypergeometric function to a polynomial of degree $n$.  Using the above notation we write the dependence of the Hypergeometric equations in terms of $\chi 's$ and $\beta's$, and therefore we have that $F_{\gamma}(\rho) =\,_{1}F_{1}\left(-\left[\frac{\beta_{\gamma}}{4\omega}-\frac{1}{2}-\frac{\vert\chi_{\gamma}\vert}{2}\right],\vert \chi_{\gamma}\vert + 1; \rho \right)$. The wavefunction can be set into the form 
\begin{eqnarray}
\Psi(\rho, \theta) = e^{-\frac{\rho}{2}} e^{ij\theta}\left(\begin{array}{c}
\rho^{\frac{|\chi_{A}|}{2}} \,_{1}F_{1}\left(-\left[\frac{\beta_{A}}{4\omega}-\frac{1}{2}-\frac{\vert\chi_{A}\vert}{2}\right],\vert \chi_{A}\vert + 1; \rho \right)\\
\rho^{\frac{|\chi_{B}|}{2}} \,_{1}F_{1}\left(-\left[\frac{\beta_{B}}{4\omega}-\frac{1}{2}-\frac{\vert\chi_{B}\vert}{2}\right],\vert \chi_{B}\vert + 1; \rho \right)\\
\rho^{\frac{|\chi_{A'}|}{2}} \,_{1}F_{1}\left(-\left[\frac{\beta_{A}}{4\omega}-\frac{1}{2}-\frac{\vert\chi_{A'}\vert}{2}\right],\vert \chi_{A'}\vert + 1; \rho \right)\\
\rho^{\frac{|\chi_{B'}|}{2}} \,_{1}F_{1}\left(-\left[\frac{\beta_{B'}}{4\omega}-\frac{1}{2}-\frac{\vert\chi_{B'}\vert}{2}\right],\vert \chi_{B'}\vert + 1; \rho \right)
\end{array}\right).
\end{eqnarray}

\begin{figure}
	\centering
		\includegraphics[width=0.50\textwidth]{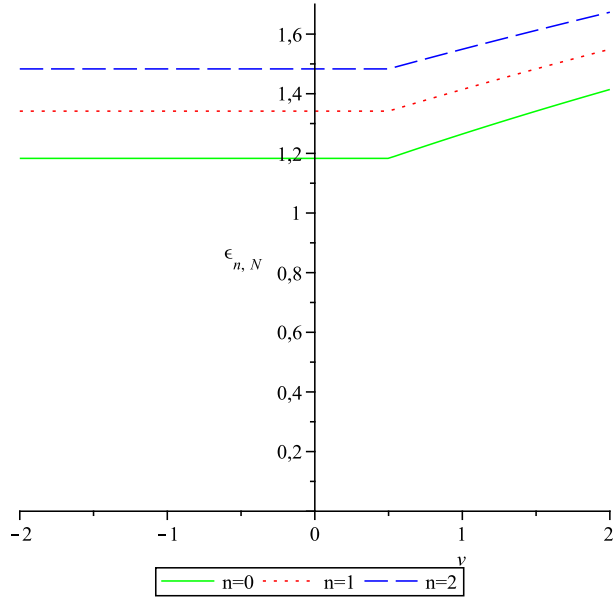}
		\caption{Plot of the energy levels as a function of $\nu$ by fixing $m=1$ and $\omega=0.1$ for the sublattice $\mathcal{A}$}
	\label{fig:fig1}
\end{figure}

\begin{figure}
\centering
		\includegraphics[width=0.50\textwidth]{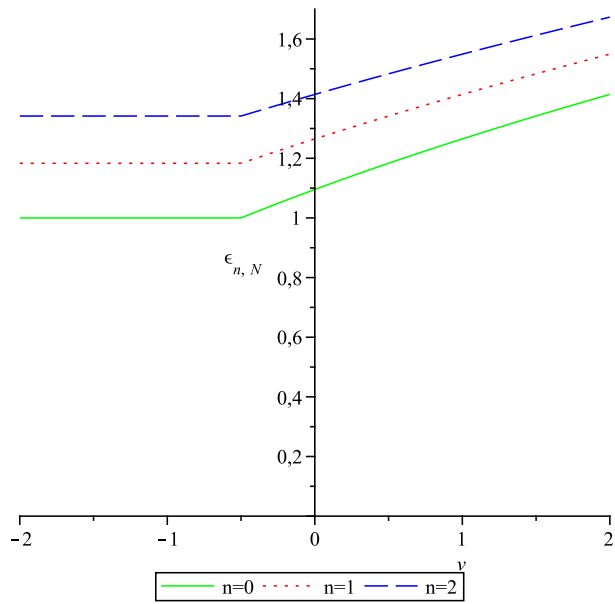}
		\caption{Plot of the energy levels as a function of $\nu$ by fixing $m=1$ and $\omega=0.1$ for the sublattice $\mathcal{B}$}
	\label{fig:fig2}
\end{figure}

The requirement of regularity conditions imposes a quantization condition on the energy levels. Explicitly, we have
\begin{eqnarray}\label{landaulevels}
\epsilon_{n,N} = \sqrt{m^{2}+4\omega n+2\omega\left[\left \vert \nu_{\tau}- \frac{\sigma_{z}}{2} \right \vert + \left(\nu_{\tau}-\frac{\sigma_{z}}{2}\right) + 1 + \sigma_{z} \right]},
\label{LandauL}
\end{eqnarray}
which are the Landau levels and recalling that $\tau$ and $\sigma_{z}$ can assume the values $\pm1$ depending on the sublattice and valley we are dealing with. Furthermore, the energy spectrum also depends on the disclination parameter $N$ which appears implicitly in $\nu_{\tau}$, see Eq.(\ref{3.7}).  To cast light on this, observe that the first two terms in the former equation describe the standard energy spectrum of Dirac fermions in the presence of a constant magnetic field, as long as the third term is a sublattice/valley-dependent one, that comes from the second-neighbors interactions and the wedge disclination. In order to interpret Eq.(\ref{LandauL}) further, it is useful to take the ``non-relativistic'' limit, $m>>2\omega=eB_0$, then:
\begin{equation}
\epsilon_{n,N}\approx \left(n+\frac{1}{2}\right)\frac{eB_{0}}{m} + m-\frac{1}{2}\frac{eB_{eff}}{m},
\end{equation}
where we have defined $B_{eff}\equiv B_{0}\left[\left \vert \nu_{\tau}- \frac{\sigma_{z}}{2} \right \vert + \left(\nu_{\tau}-\frac{\sigma_{z}}{2}\right) + \sigma_{z} \right]$. Within this approximation, the energy spectrum might be separated into two parts: the Landau level spectrum for a constant magnetic field added by a zero point energy, as a result, even for the ground state, $n=0$, there exists a non-trivial Landau level \cite{Jackiw} which, in our case, relies on $m$, $B_{eff}$. The degeneracy in each Landau level, in particular, has now been raised. To the best of our knowledge, let us go back to the full expression of the energy spectrum. We illustrate the plot of the energy level as a function of the parameter $\nu$ for the sublattice $\mathcal{A}$, Fig.(\ref{fig:fig1}), and $\mathcal{B}$, Fig.(\ref{fig:fig2}). Observe from Fig.(\ref{fig:fig1}) that the new effects of the second neighbors interactions and the wedge disclination are triggered within the range $\nu>\frac{1}{2}$; otherwise, these effects are turned off and then we recovered the Landau levels of Dirac fermions in the presence of a constant magnetic field as one has already been deduced by taking the non-relativist limit. We check a similar effect for the sublattice $\mathcal{B}$ displayed in Fig.(\ref{fig:fig2}), though, the new effects are now turned on within the range $\nu>-\frac{1}{2}$.   

\begin{figure}
\centering
		\includegraphics[width=0.50\textwidth]{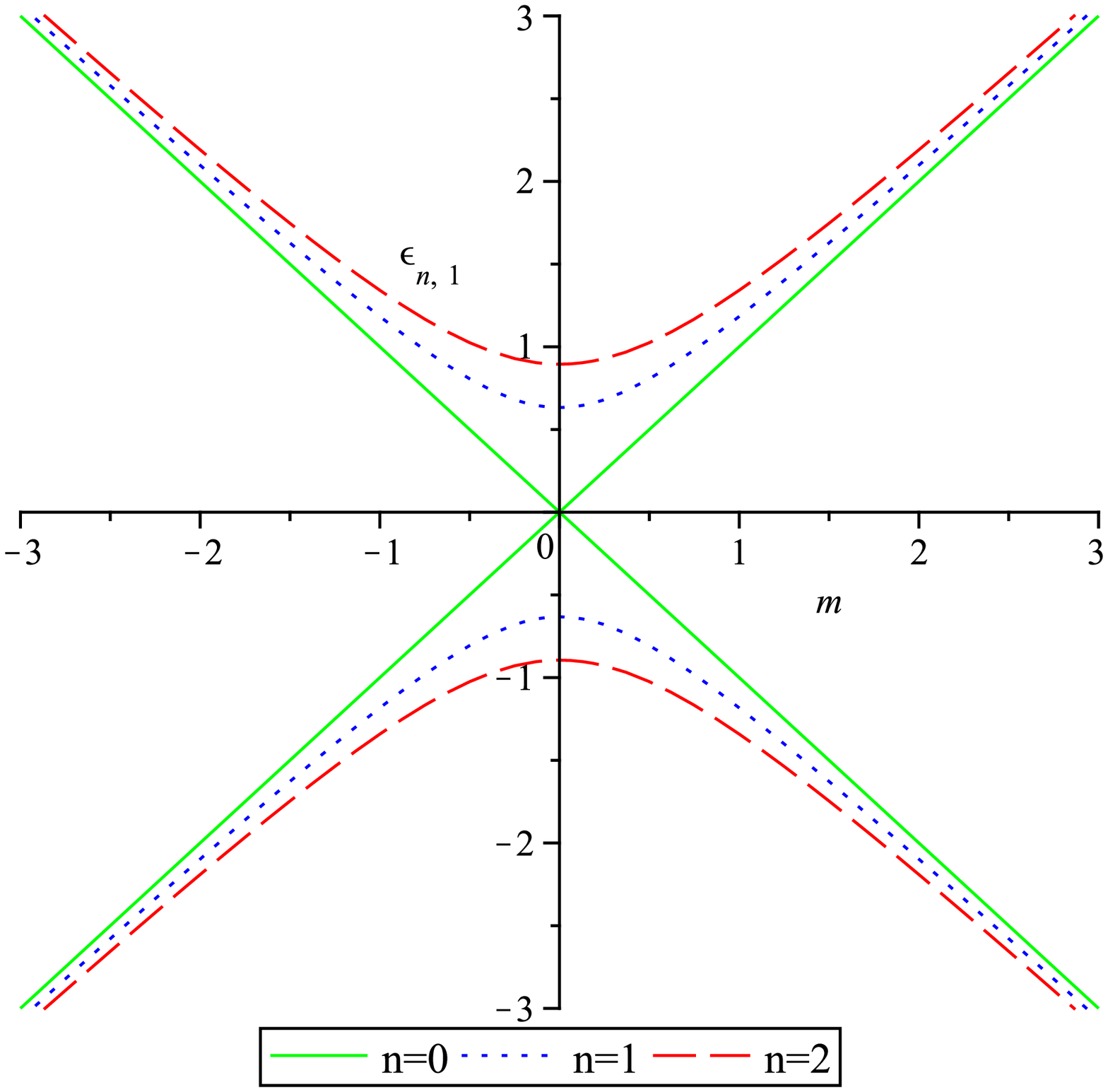}
		\caption{Plot of the energy levels as a function of $m$ by fixing $\omega=0.1$ for the sublattice $\mathcal{B}$ and valley $K^{\prime}$ }
	\label{fig:fig3}
\end{figure}

\begin{figure}
\centering
		\includegraphics[width=0.50\textwidth]{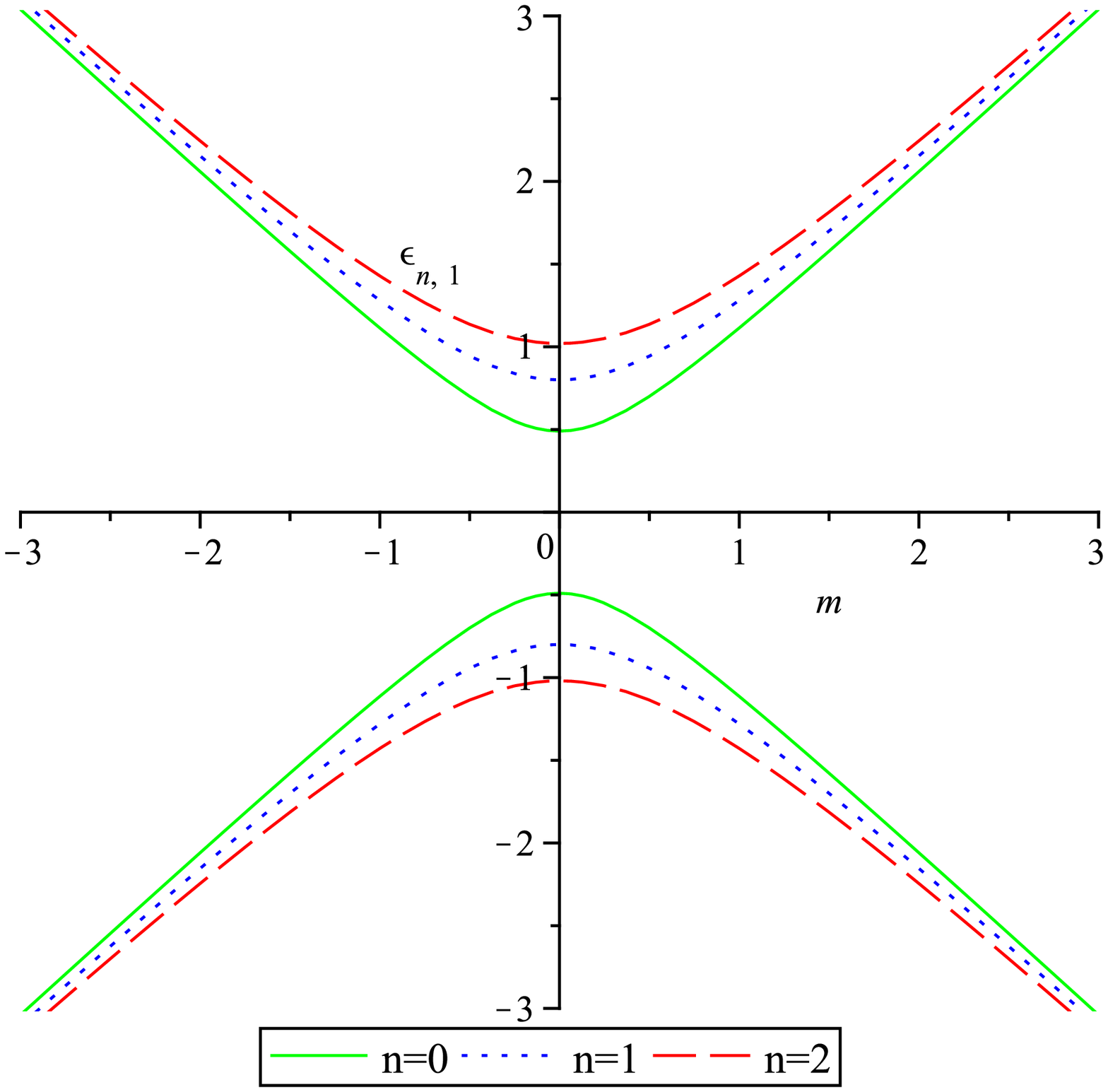}
		\caption{Plot of the energy levels as a function of $m$ by fixing $\omega=0.1$ for the sublattice $\mathcal{B}$ and valley $K$ }
	\label{fig:fig4}
\end{figure}

\begin{figure}
\centering
		\includegraphics[width=0.50\textwidth]{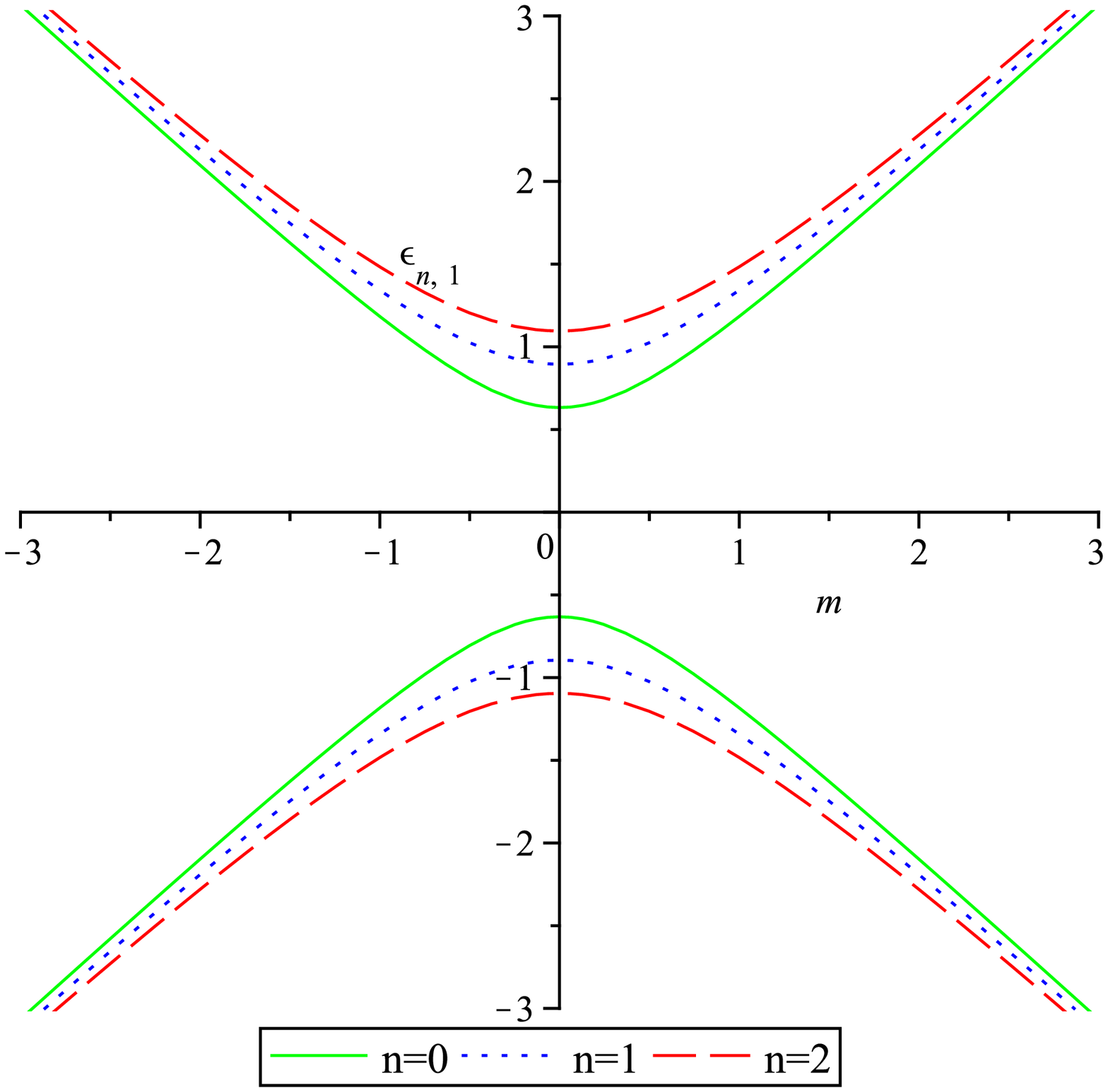}
		\caption{Plot of the energy levels as a function of $m$ by fixing $\omega=0.1$ for the sublattice $\mathcal{A}$ and valley $K^{\prime}$ }
	\label{fig:fig5}
\end{figure}

\begin{figure}
\centering
		\includegraphics[width=0.50\textwidth]{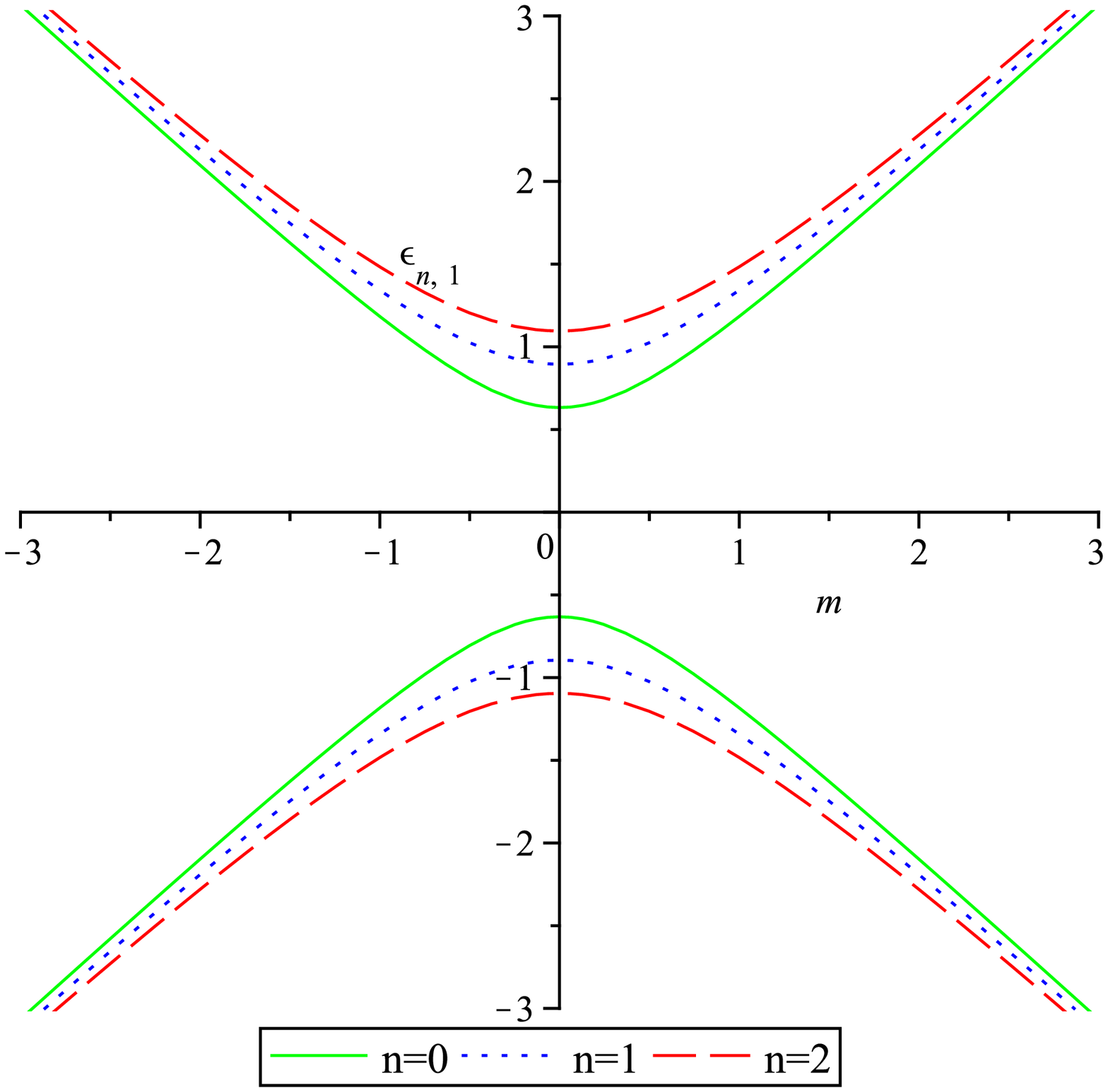}
		\caption{Plot of the energy levels as a function of $m$ by fixing $\omega=0.1$ for the sublattice $\mathcal{A}$ and valley $K$ }
	\label{fig:fig6}
\end{figure}

Now we will examine the energy spectrum in terms of the mass. In order to do so, let us pick $j=\frac{1}{2}, \frac{\Phi}{\Phi_0}=-\frac{2}{3}$ and $N=1$ (it corresponds to a pentagon defect). Interestingly, the Fig.(\ref{fig:fig3}) exhibits a gapless graphene layer, thus the system behaviors as a metal at the valley $K^{\prime}$ and sublattice $\mathcal{B}$. On the other hand, the Fig.(\ref{fig:fig4}) displays a gapped graphene layer, in this case, the system behaviors as an insulator at the valley $K$ and sublattice $\mathcal{B}$. Figs.(\ref{fig:fig5}) and (\ref{fig:fig6}) show the same behavior (insulator) for both valleys.

\section{Summary and Conclusions}\label{End}
In this work we have considered a particular modified Kane-Mele Hamiltonian (with trivial SOC) in the presence of a wedge disclination plus a mass term. First, by solving the eigenvalue equation for the Hamiltonian, we have been able to find the quantized energy spectrum by imposing regularity conditions for the hypergeometric functions. Once the energy spectrum is known, we have seen that they can be interpreted as: the standard Landau levels plus a zero point energy which in turn depends on the disclination parameter $\alpha_{N}$ or $N$. As a straightforward consequence, we have found that the degeneracy of the Landau levels is lifted within the ranges $\nu>\frac{1}{2}$ and  $\nu>-\frac{1}{2}$ at the sublattice $\mathcal{A}$ and $\mathcal{B}$, respectively.

We have also realized the existence of gapless and gapped states depending on the valley we are dealing with. This is the main effect observed by the presence of wedge disclinations in the energy spectrum.

{\bf Acknowledgements}

G.Q.G would like to acknowledge to FAPESQ/PB and CNPq for financial support. C.F.  thanks to CAPES and  CNPQ for financial support. P.J.P. would like to acknowledge the Brazilian agency CAPES for financial support.


\end{document}